\documentclass{emulateapj}
\usepackage{natbib}
\usepackage{amssymb}
\usepackage{setspace}
\usepackage{graphicx}
\usepackage{rotating}
\usepackage{color}
\submitted{Accepted for Publication in the Astrophysical Journal}

\def\CIVdblt{{\rm C~}\kern 0.1em{\sc iv}~$\lambda\lambda 1548, 1550$}
\def\MgIIdblt{{\rm Mg~}\kern 0.1em{\sc ii}~$\lambda\lambda 2796, 2803$}
\def\NVdblt{{\rm N}\kern 0.1em{\sc v}~$\lambda\lambda 1238, 1242$}
\def\OVIdblt{{\rm O}\kern 0.1em{\sc vi}~$ 1031, 1037$}
\def\SiIVdblt{{\rm Si~}\kern 0.1em{\sc iv}~$\lambda\lambda1394, 1403$}
\def\AlIIIdblt{{\rm Al~}\kern 0.1em{\sc iii}~$\lambda\lambda1855,1863$}
\def\FeIIdblt{{\rm Fe~}\kern 0.1em{\sc ii}~$\lambda\lambda 2383, 2600$}
\def\NeVIIIdblt{{\rm Ne~}\kern 0.1em{\sc viii}~$ 770, 780$}

\def\NeVIII{\hbox{{\rm Ne~}\kern 0.1em{\sc viii}}}
\def\OI{\hbox{{\rm O~}\kern 0.1em{\sc i}}}
\def\OII{\hbox{{\rm O~}\kern 0.1em{\sc ii}}}
\def\OIII{\hbox{{\rm O~}\kern 0.1em{\sc iii}}}
\def\OIV{\hbox{{\rm O~}\kern 0.1em{\sc iv}}}
\def\OV{\hbox{{\rm O~}\kern 0.1em{\sc v}}}
\def\OVI{\hbox{{\rm O~}\kern 0.1em{\sc vi}}}
\def\OVII{\hbox{{\rm O~}\kern 0.1em{\sc vii}}}
\def\OVIII{\hbox{{\rm O~}\kern 0.1em{\sc viii}}}
\def\NIII{\hbox{{\rm N~}\kern 0.1em{\sc iii}}}
\def\NIV{\hbox{{\rm N~}\kern 0.1em{\sc iv}}}
\def\NVII{\hbox{{\rm N~}\kern 0.1em{\sc vii}}}
\def\CIII{\hbox{{\rm C~}\kern 0.1em{\sc iii}}}
\def\SiIII{\hbox{{\rm Si~}\kern 0.1em{\sc iii}}}
\def\SVI{\hbox{{\rm S~}\kern 0.1em{\sc vi}}}
\def\NeIX{\hbox{{\rm Ne~}\kern 0.1em{\sc ix}}}

\def\AlII{\hbox{{\rm Al~}\kern 0.1em{\sc ii}}}
\def\AlIII{\hbox{{\rm Al~}\kern 0.1em{\sc iii}}}
\def\CaI{\hbox{{\rm Ca}\kern 0.1em{\sc i}}}
\def\CaII{\hbox{{\rm Ca}\kern 0.1em{\sc ii}}}
\def\CrII{\hbox{{\rm Cr}\kern 0.1em{\sc ii}}}
\def\CII{\hbox{{\rm C~}\kern 0.1em{\sc ii}}}
\def\CIII{\hbox{{\rm C~}\kern 0.1em{\sc iii}}}
\def\CIV{\hbox{{\rm C~}\kern 0.1em{\sc iv}}}
\def\CV{\hbox{{\rm C}\kern 0.1em{\sc v}}}
\def\H{\hbox{{\rm H}}}
\def\HI{\hbox{{\rm H~}\kern 0.1em{\sc i}}}
\def\HII{\hbox{{\rm H~}\kern 0.1em{\sc ii}}}
\def\Lya{\hbox{{\rm Ly}\kern 0.1em$\alpha$}}
\def\Lyb{\hbox{{\rm Ly}\kern 0.1em$\beta$}}
\def\Lyg{\hbox{{\rm Ly}\kern 0.1em$\gamma$}}
\def\Lyth{\hbox{{\rm Ly}\kern 0.1em$\theta$}}
\def\Lyfive{\hbox{{\rm Ly}\kern 0.1em$5$}}
\def\Lysix{\hbox{{\rm Ly}\kern 0.1em$6$}}
\def\Lyseven{\hbox{{\rm Ly}\kern 0.1em$7$}}
\def\Lyeight{\hbox{{\rm Ly}\kern 0.1em$8$}}
\def\Lynine{\hbox{{\rm Ly}\kern 0.1em$9$}}
\def\Lyten{\hbox{{\rm Ly}\kern 0.1em$10$}}
\def\HeI{\hbox{{\rm He}\kern 0.1em{\sc i}}}
\def\HeII{\hbox{{\rm He}\kern 0.1em{\sc ii}}}
\def\FeI{\hbox{{\rm Fe~}\kern 0.1em{\sc i}}}
\def\FeII{\hbox{{\rm Fe~}\kern 0.1em{\sc ii}}}
\def\FeIII{\hbox{{\rm Fe~}\kern 0.1em{\sc iii}}}
\def\MnII{\hbox{{\rm Mn}\kern 0.1em{\sc ii}}}
\def\MgI{\hbox{{\rm Mg~}\kern 0.1em{\sc i}}}
\def\MgII{\hbox{{\rm Mg~}\kern 0.1em{\sc ii}}}
\def\MgIII{\hbox{{\rm Mg~}\kern 0.1em{\sc iii}}}
\def\MgIV{\hbox{{\rm Mg~}\kern 0.1em{\sc iv}}}
\def\NaI{\hbox{{\rm Na}\kern 0.1em{\sc i}}}
\def\NV{\hbox{{\rm N}\kern 0.1em{\sc v}}}
\def\NII{\hbox{{\rm N}\kern 0.1em{\sc ii}}}
\def\NIII{\hbox{{\rm N}\kern 0.1em{\sc iii}}}
\def\OVI{\hbox{{\rm O}\kern 0.1em{\sc vi}}}
\def\SiII{\hbox{{\rm Si~}\kern 0.1em{\sc ii}}}
\def\SiIII{\hbox{{\rm Si~}\kern 0.1em{\sc iii}}}
\def\SiIV{\hbox{{\rm Si~}\kern 0.1em{\sc iv}}}
\def\SII{\hbox{{\rm S}\kern 0.1em{\sc ii}}}
\def\SIII{\hbox{{\rm S}\kern 0.1em{\sc iii}}}
\def\SIV{\hbox{{\rm S}\kern 0.1em{\sc iv}}}
\def\TiII{\hbox{{\rm Ti}\kern 0.1em{\sc ii}}}
\def\ZnII{\hbox{{\rm Zn}\kern 0.1em{\sc ii}}}
\newcommand{\kms}{\hbox{km~s$^{-1}$}}
\newcommand{\cmsq}{\hbox{cm$^{-2}$}}
\newcommand{\cc}{\hbox{cm$^{-3}$}}
\def\kms{\hbox{km~s$^{-1}$}}
\def\cmsq{\hbox{cm$^{-2}$}}
\def\cc{\hbox{cm$^{-3}$}}

\newcommand {\apgt} {\ {\raise-.5ex\hbox{$\buildrel>\over\sim$}}\ }
\newcommand {\aplt} {\ {\raise-.5ex\hbox{$\buildrel<\over\sim$}}\ } 

\begin{document}

\title{Cosmic Origins Spectrograph Observations of \\ Warm Intervening Gas at $z \sim 0.325$ Towards 3C~$263$ \altaffilmark{1}}

\author{Anand Narayanan\altaffilmark{2}, Blair D. Savage, \& Bart P. Wakker\altaffilmark{3}}

\altaffiltext{1}{Based on observations with the NASA/ESA {\it Hubble Space Telescope}, obtained at the Space Telescope Science Institute, which is operated by the Association of Universities for Research in Astronomy, Inc., under NASA contract NAS 05-26555, and the NASA-CNES/ESA {\it Far Ultraviolet Spectroscopic Explorer} mission, operated by the Johns Hopkins University, supported by NASA contract NAS 05-32985.}
\altaffiltext{2}{Indian Institute of Space Science \& Technology, Thiruvananthapuram 695 547, Kerala, INDIA. Email: anand@iist.ac.in}
\altaffiltext{3}{Department of Astronomy, The University of Wisconsin-Madison, 5
534 Sterling Hall, 475 N. Charter Street, Madison WI 53706-1582, USA, Email: savage, wakker@astro.wisc.edu}
\subjectheadings{galaxies: halos, intergalactic medium, quasars: absorption lines, quasars: individual: 3C~263, ultraviolet: general}

\begin{abstract}

We present $HST$/COS high $S/N$ observations of the $z = 0.32566$ multi-phase absorber towards 3C~263. The COS data shows absorption from {\HI} (Ly-$\alpha$ to Ly-$\theta$), {\OVI}, {\CIII}, {\NIII}, {\SiIII} and {\CII}. The {\NeVIII} in this absorber is detected in the $FUSE$ spectrum along with {\OIII}, {\OIV}, and {\NIV}. The low and intermediate ions are kinematically aligned with each other and {\HI} and display narrow line widths of $b \sim 6 - 8$~{\kms}. The {\OVIdblt}~{\AA} lines are kinematically offset by $\Delta v \sim 12$~{\kms} from the low ions and are a factor of $\sim 4$ broader. All metal ions except {\OVI} and {\NeVIII} are consistent with an origin in gas photoionized by the extragalactic background radation. The bulk of the observed {\HI} is also traced by this photoionized medium. The metallicity in this gas phase is Z $ \apgt 0.15$~Z$_\odot$ with carbon having near-solar abundances. The {\OVI} and {\NeVIII} favor an origin in collisionally ionized gas at $T = 5.2 \times 10^5$~K. The {\HI} absorption associated with this warm absorber is a BLA marginally detected in the COS spectrum. This warm gas phase has a metallicity of [X/H] $\sim -0.12$~dex, and a total hydrogen column density of $N(\H) \sim 3 \times 10^{19}$~{\cmsq} which is $\sim 2$~dex higher than what is traced by the photoionized gas. Simultaneous detection of {\OVI}, {\NeVIII} and BLAs in an absorber can be a strong diagnostic of gas with $T \sim 10^5 - 10^6$~K corresponding to the {\it warm} phase of the WHIM or shock-heated gas in the extended halos of galaxies.

\end{abstract}

\section{Introduction}

Throughout cosmic history, the intergalactic medium (IGM) and the gaseous envelopes surrounding galaxies have retained more baryons compared to the gas settled into galaxies. In the $z < 0.5$ universe, $> 50$\% of the baryons are predicted to be in the form of low density ($n_{\H} \sim 10^{-5}$~{\cc}) intergalactic gas at temperatures of $T \sim 10^5 - 10^7$~K and moderate overdensities of $\rho/\bar{\rho} \sim 20$. These baryons, which were once part of the cool ($T \aplt 10^4$~K) photoionized IGM probed by the {\Lya} forest at $z \apgt 3$, were heated through gravitational shocks during the formation of large scale structures \citep{cen99, dave01, valageas02}. The temperatures imply that this warm-hot intergalactic medium (WHIM) gas has a high degree of ionization. Physical conditions similar to the WHIM can also exist in gas in the extended halos of galaxies. Cosmological simulations predict that massive halos ($> 10^{12}$~M$_\odot$) acquire most of their gas through the {\it hot-mode} of accretion, where the infalling intergalactic gas is shock-heated to the virial temperatures of $T \apgt 10^6$~K \citep{keres05, vande11}. This infalling gas may circulate within the halo for a long time before it can radiatively cool and flow into the disc. Other galactic-scale processes such as supernova driven flows and tidal interactions/mergers can also increase the temperature and ionization levels of gas in regions close to galaxies. 

Observations of warm gas with $T \sim 10^5 - 10^6$~K in the low-$z$ universe has been accomplished primarily through quasar absorption line spectroscopy in the far-UV of highly ionized metals, particularly {\OVIdblt}~{\AA} lines \citep{tripp00, richter04, danforth05, stocke06, danforth08, tripp08} and more recently {\NeVIIIdblt}~{\AA} \citep{savage05, narayanan09, narayanan11, meiring12}. The fractional abundances of these ions peak in the interval $T \sim (2 - 7) \times 10^5$~K when ionizations are controlled by ion-electron collisions. Limitations in accessing the FUV wavelengths of the {\NeVIII} resonant transitions and the relatively lower cosmic abundance of neon compared to oxygen has resulted in fewer {\NeVIII} findings compared to {\OVI}. On the other hand, the detection of {\NeVIII} has the advantage that it requires the presence of collisionaly ionized warm gas. In the case of {\OVI} it is not straightforward to distinguish a photoionization origin from collisional ionization in warm gas. 

A strong correlation between the incidence of {\OVI} absorption and galaxies has been emerging from recent absorber-galaxy surveys. The {\OVI} absorption at low-$z$ seems to be preferentially selecting circumgalactic environments of typical star forming galaxies \citep{savage02, fox04, stocke06, wakker09, tumlinson11b, meiring12}. The covering fraction of {\OVI} absorbing gas is estimated to be $\apgt 65$\% around emission line galaxies \citep{chen09}, comparable to the covering fraction of high-velocity {\OVI} around the Milky Way \citep{sembach03}. In the case of the Milky Way, the halo {\OVI} absorption is known to be from multi-phase high velocity clouds where the {\OVI} is produced in $T \sim (2 - 5) \times 10^5$~K transition temperature plasma at the interface layers between the $T \aplt 10^4$~K neutral interstellar gas in HVCs and the $T \sim 10^6$~K coronal halo of the Galaxy \citep{sembach03, fox06}. As analogs of Galactic HVCs, some fraction of the population of extragalactic warm absorbers could be tracing halo gas associated with galactic-scale processes such as outflows from star formation \citep{heckman01}, accretion of WHIM gas from the nearby intergalactic filaments \citep{narayanan10b, tumlinson11a} and/or tidal streams from interactions and mergers with companion galaxies similar to the Magellanic Stream around the Milky Way, with which {\OVI} absorption is clearly associated \citep{sembach03}. 

In the few known {\NeVIII} absorbers, associated {\OVI} has always been detected, and both ions are found to be tracing warm gas with $T \apgt 10^{5}$~K \citep{savage05, narayanan11}. Simultaneous detection of {\NeVIII} and {\OVI} can be a strong diagnostic on the temperature of the gas. The warm temperature and high levels of collisional ionization would result in thermally broadened ($b(\HI) > 40$~{\kms}) and often shallow {\Lya} features \citep{richter06, danforth10, narayanan10b}. Detection of absorption from highly ionized species and broad-{\Lya} absorbers (BLAs) in the low-$z$ universe can be used as probes of the WHIM or halo gas transiting between galaxies and the surrounding IGM. 

This paper adds important FUV $HST$/Cosmic Origins Spectrograph (COS) data on the $z = 0.32566$ {\NeVIII} absorber previously detected with $FUSE$ data \citep{narayanan09}. The detection of {\NeVIII} indicated the presence of $T \apgt 10^5$~K collisionally ionized gas, but the relatively few lines covered in the $FUSE$ spectrum ({\OIII}, {\OIV} and {\NIV}) were inadequate to constrain the ionization conditions and metallicity in this absorber. The COS spectra are obtained at high $S/N$ and provide coverage of a large number of absorption lines, most importantly {\OVI} and {\HI}. In Sec 2 and 3 we describe the COS observations and $FUSE$ observations for this sight line. The line detections and the multiphase properties of the absorber are discussed in detail in Sec 4, followed by predictions from photoionization and collisional ionization models in Sec 5. 

\section{COS Observations of 3C~263}

COS observations of 3C 263 were carried out as part of the $HST$ Cycle 19 GTO Program ID 11541 (PI Green). The capabilities of COS are described in detail by \citet{green12}, \citet{froning09} and the on-orbit performance of the instrument is discussed by \citet{osterman11}. The 3C~263 observations consisted of FUV spectra obtained at intermediate resolutions (FWHM $\sim 17$~{\kms}) using the G130M and G160M COS gratings with total exposure times of 15.4 ks and 18.0 ks respectively. The details on the individual exposures are given in Table 1. Different grating central wavelength settings were used for the separate exposures. For different grating central wavelength settings the dispersed light for a particular wavelength falls on a different region of the detector. This helps to reduce the amplitude of detector fixed pattern noise in the final coadded spectrum. The setup also allows for the coverage of the $\sim 9$~{\AA} wavelength gap introduced by the separation between the two segments of the COS detector. The data were extracted using the STScI CalCOS v2011.1a pipeline. The separate one dimensional spectra were coadded in flux units weighted by their respective exposure time using the routine developed by Charles Danforth and the COS GTO team\footnote{http://casa.colorado.edu/$\sim$danforth/science/cos/costools.html}. The final coadded spectrum has a wavelength coverage of $1150 - 1750$~{\AA} with a S/N $\sim 18 - 35$ per 17~{\kms} resolution element. The pipeline reduced data are however sampled at 6 detector pixels per resolution element and throughout we display the COS spectra with this sampling. The COS wavelength calibration is accurate to $\sim 15$~{\kms} but has been improved to $\sim 8$~{\kms} through cross referencing of prominent IGM and ISM absorption lines.  

\section{$FUSE$ Observations of 3C~263}

The 3C~263 was observed by $FUSE$ \citep{moos00, sahnow00} for a total of 260 ks under various observing programs. The spectra were processed using the CALFUSE (ver 2.4) data reduction pipeline software. $FUSE$ covers the wavelength range from $912$~{\AA} to $1185$~{\AA} sampling the spectrum at a resolution of $\sim 20$~{\kms} (FWHM). The $S/N$ of the spectrum at $\lambda > 1000$~{\AA} is $\sim 10 - 15$ per 17~{\kms} bin size. The procedures adopted in the coaddition of the spectrum and for correcting the zero-point velocity offset errors are similar to the detailed description given in \citet{wakker03}. More details can be found in \citet{narayanan09}. 

\section{Multi-Phase Nature of $z = 0.32566$ Absorber}

In Figures 1a - 1b, we display the $z = 0.32566$ centered system plot with prominent metal lines and {\HI} detected by both COS and $FUSE$.  To be consistent with the previous $FUSE$ analysis of this absorber, we adopt the same system redshift as given in \citet{narayanan09}. The continuum normalization was done by fitting low-order polynomials to the region around each absorption feature. The metal lines with $> 3 \sigma$ detection by COS for this absorber are {\CII}~$903.96$~{\AA}, {\CII}~$903.62$~{\AA}, {\CII}~$1036$~{\AA}, {\CIII}~$977$~{\AA}, {\NIII}~$989$~{\AA}, {\SiIII}~$1206$~{\AA} and {\OVIdblt}~{\AA}. In the system plot we also show non-detections of lines corresponding to {\NII} and {\SiII}. In $FUSE$, {\OIII}~$832$~{\AA}, {\OIV}~$787$~{\AA}, {\NIV}~$765$~{\AA} and {\NeVIII}~$770$~{\AA} are seen along with a non-detection of {\OII}~$834$~{\AA}. 

Measurements on all the lines were carried out using the apparent optical depth (AOD) technique of \citet{savage91}. The AOD measurements on the COS data are listed in Table 2.  The column density $N$, Doppler parameter $b$ and velocity $v$ of the line components were also determined through Voigt profile modeling of the lines using the \citet{fitzpatrick97} routine. While performing the fits, the model profiles were convolved with the emperically determined line spread functions of \citet{kriss11} for the redshifted wavelength of each line. The fitting results are given in Table 3.  

The COS spectrum shows strong absorption from {\CIII}~$977$~{\AA} with at least two components at velocities of -23~{\kms} and +12~{\kms}. Coincident with the positive velocity component are also seen absorption from {\NIII}, {\SiIII}, {\CII} and {\HI}. For the negative velocity component, these metal ions are non-detections in the COS spectrum down to the 3$\sigma$ significance level. The difference in the corresponding line strengths between the two components is indicative of ionization or metallicity gradients within the absorber. The {\CIII}~$977$~{\AA} being a strong line ($f_\mathrm{osc} = 0.762$) is susceptible to saturation unresolved at the FWHM $\sim 17$~{\kms} of the COS data. Strong saturation will result in the AOD measurements underestimating the column densities in the line cores. We find a difference of $\sim 0.32$~dex in the column densities between the two measurements. The profile fit to {\CIII} yields a $b(\CIII) = 9~{\pm}~3$~{\kms} for the positive velocity component. The narrower $b$-values for {\CII} and {\SiIII} indicate that the true dispersion in the low ionization gas could be smaller than what we measure for {\CIII}. Instrumental broadening of narrow spectral lines would result in the column density getting underestimated so as to preserve the equivalent width. Lowering the $b(\CIII)$ from its measured value of $9$~{\kms} to $6$~{\kms} (the $1\sigma$ lower limit) yields a column density which is 0.7~dex higher than the free-fit value of 13.67~dex. This shows that the degree of unresolved saturation is significant if the $b(\CIII)$ is narrower than what we measure. To account for this possibility, we increase the $+1\sigma$ uncertainty to 0.7 dex in the profile fit $N$-value for the positive velocity component of the {\CIII}~$977$~{\AA} line. 

The {\SiIII}~1207~{\AA} line is covered by the G160M grating of the COS spectrum. The positive velocity component is very narrow with a measured $b (\SiIII) = 5~{\pm}~3$~{\kms}. The profile fit column density is $\sim 0.3$~dex larger than the apparent column density, but within the combined $\sim 1\sigma$ errors associated with the measurements. At $v \sim -20$~{\kms}, weak absorption is detected at $\sim 4 \sigma$ significance which is consistent with being {\SiIII} corresponding to the negative component seen in {\CIII} and {\HI}. The {\CII} multiplet transitions at $903.6235$~{\AA} ($f_{osc} = 0.168$), $903.9616$~{\AA} ($f_{osc} = 0.336$) and $1036.3367$ ($f_{osc} = 0.123$) are detected with $> 3\sigma$ significance. In the $N_a(v)$ comparsion in Figure 3, we find the apparent column density profile of the stronger {\CII} line is lower than the weaker lines. The {\CII} line is possibly stronger and narrower than the observed profile but has been blurred by the instrumental spread function. The $N_a(v)$ integrated column density obtained is therefore only a lower limit. The difference in $N_a$ values between the weaker ({\CII}~$1036$~{\AA}) and the stronger ({\CII}~$903.9616$~{\AA}) transitions is $0.11$~dex. This difference can be used to compensate for the instrumental broadening, as described in \citet{savage91}. The corrected apparent column density measurement of $\log N_a(\CII) = \log N_a(\CII~1036) + \Delta\log N_a = 13.29~{\pm}~0.11$~dex is consistent with the column density obtained from simultaneous profile fitting of the three lines. The $N_a(v)$ comparison in Figure 3 also shows small excess absorption in the {\CII}~$903.9616$~{\AA} line between $30 < v < 75$~{\kms} possibly due to line contamination. The separate profile fit on this line yields a larger $b$ value compared to the two weaker {\CII} transitions. We obtain $\log N(\CII) = 13.26~{\pm}~0.03$~dex and $b(\CII) = 8~{\pm}~2$~{\kms} from simultanesouly fitting the three {\CII} lines. A lower $b(\CII) = 6$~{\kms} would result in $\log N(\CII) = 13.28$~{\kms} which is within the statistical uncertainty in the measurement. 

The COS spectrum shows absorption in {\HI} from Ly-$\alpha$ to Ly-$\theta$. The Ly-$\delta$ ({\HI}~949~{\AA}) line suffers contamination from Galactic {\SiII}~1260~{\AA} and possibly also from {\SII}~1260~{\AA}. The {\Lya} and {\Lyb} lines have saturated profiles. The column density and the component structure of {\HI} are best constrained in the higher order Lyman lines with kinematic sub-structure evident in the {\HI}~$938$~{\AA} and {\HI}~$926$~{\AA} profiles. Due to the effect of random noise, the corresponding component structures are not conspicuous in all higher order Lyman lines. We simultaneously fitted the Lyman series lines (except Ly$\delta$ which is affected by contamination) by keeping $v$, $b$ and $N$ as free parameters. The best-fit model gave two components of roughly equal strength contributing to the core {\HI} absorption (see Table 3). The $b \sim 17$~{\kms} for either component is consistent with temperature for photoionized gas. A comparison of the $N_a(v)$ profiles of {\HI}~938 and {\HI}~$926$~{\AA} lines with {\CIII}~977~{\AA} show similar two component velocity structure (see Figure 3). This kinematic coincidence indicates that the bulk of the {\HI} is contributed by the same gas phase as {\CIII}. 

The wings of the {\Lya} profile show the possible presence of a very broad component superimposed on the saturated {\HI} core and spread over the velocity interval [$-v, +v$] = [$-130, +130$]~{\kms}. This excess absorption could also be due to the presence of additional narrow kinematic sub-structures. The resolution of COS is not adequate to rule out this additional complexity in {\HI} kinematics. The total column density of {\HI} obtained from the free-fit to the Lyman transitions however does not account for this extra absorption in the far wings of the {\Lya} profile. We discuss more about this possible broad-{\Lya} feautre in Sec 7. 
 
The {\OVI} is a strong absorption seen in both members of the doublet. The $N_a(v)$ profiles for the {\OVIdblt}~{\AA} lines, shown in Figure 3, are in good agreement with each other suggesting little contamination or unresolved saturation. The doublet lines were simultaneously fitted with a single component. The profiles do not shown evidence for any kinematic substructure. The parameters of the profile fit are given in Table 3. The central velocity of the {\OVI} absorption is distinct from the velocities of either component seen in {\CIII} or the core absorption in {\HI}. Also, the best-fit $b$ parameter for the {\OVI} line is a factor of $\sim 4$ broader than the {\CIII} or {\HI} line widths, which  suggests that the two ions are tracing separate gas phases. 

In Table 4, we report the equivalent widths and the total apparent column densities for the $> 3\sigma$ transitions seen in the $FUSE$ spectrum. The $FUSE$ spectrum has a factor of $\gtrsim 3$ lower $S/N$ compared to COS. At the low $S/N$, the component structure is not evident in the {\OIII}~$833$~{\AA}, {\OIV}~$788$~{\AA}, {\NIV}~$765$~{\AA} or {\NeVIII}~770~{\AA} lines. The integration range of the $N_a(v)$ profiles for the intermediate ion transitions were therefore broken into two regions of [$-75,-10$]~{\kms} and [$-10, 75$]~{\kms} to match with the component structure seen in {\CIII}, {\NIII}, {\SiIII}, {\CII} and {\HI} in the COS spectrum. To compensate for our lack of independent information on the component structure in the low $S/N$ data, we double the logarithmic errors in the $FUSE$ $N_a$ measurements. 

\section{Ionization Modeling of the Multiphase Absorber}

\subsection{Photoionization of the Low \& Intermediate Ions}

The large difference in the $b$-parameter and the kinematic offset of $|\Delta v| \sim 12$~{\kms} between {\OVI} and the low/intermediate metal ions and the {\HI} core points to the presence of multiple ionization phases in the absorber. For modeling the physical conditions in the absorber, we consider the two possible scenarios of ionization by EUV radiation and ionization from ion-electron collisions. The {\HI} and {\CIII} lines show absorption in two kinematically distinct components. The column density ratios of the various ions in these two components suggests different ionization conditions in the two clouds. However constraints are adequate only for the positive velocity component, and hence we only model that component. Using the photoionization code Cloudy (version C08.00; \citet{ferland98}) we solve for the models that best reproduce the observed column density ratios. We have assumed an ionizing background radiation field with contributions from both AGNs and star forming galaxies as modeled by \citet{haardt01}. 

In the positive velocity cloud, the constraint on density comes from the column density ratios of C, N and O in their adjacent ionization levels, $\log~N(\CII)/N(\CIII) \sim -0.41$, $\log~N(\NIII)/N(\NIV) \sim 0.65$ and $\log~N(\OIII)/N(\OIV) \sim 0.15$, along with upper limits of $\log~N(\NII)/N(\NIII) \aplt -0.75$, $\log~N(\OII)/N(\OIII) \aplt -0.20$, and $\log~N(\SiII)/N(\SiIII) \aplt 0.04$. By fixing the {\HI} column density in this component to the measured value of $\log~N(\HI) = 15.20$~dex, we ran a series of Cloudy models for a range of ionization parameters. The ionic column densities predicted by the photoionization models are shown in Figure 4. The models are consistent with the observed column density ratios within the narrow interval of $-2.6 \leq \log~U \leq -2.1$. This corresponds to a density range of $n_{\H} \sim (0.4 - 2)~\times~10^{-3}$~{\cc}. The single phase photoionization model which best fits the low and intermediate ions at $\log~U \sim -2.2$ predicts $n_{\H} \sim 5 \times 10^{-4}$~{\cc}, total hydrogen column density of $N(\H) \sim 2 \times 10^{18}$~{\cmsq}, $T \sim 1.4 \times 10^4$~K, pressure of $p/K \sim 16$~K~{\cc}, and a path length of $\sim 1.6$~kpc through the absorber.  

For the given $N(\HI)$ value, $N(\OIII)$ and $N(\OIV)$ are simultaneously predicted at $\log~U \sim -2.2$ when the oxygen abundance is [O/H] $= -0.8~{\pm}~0.1$~dex. The uncertainty comes from the $1\sigma$ errors in the column density measurements. To reproduce the other low and intermediate ionic column densities from the same phase, the abundances have to be [C/H] $ = 0~{\pm}~0.04$~dex, [N/H] $= -0.3~{\pm}~0.1$, and [Si/H] $= -0.6~{\pm}~0.3$~dex. The higher abundance of carbon is required to explain the observed $N(\CII)$. The model prediction for $N(\CII)$ has its peak at $\log~U \sim -2.5$ and does not vary much in the interval $-3.2 \aplt \log~U \aplt -1.8$, which sets a robust constrain on the carbon abundance. The estimated abundances are subjected to larger systematic uncertainties at the level of $\sim 0.4$~dex because of ambiguities in the shape and intensity of the ionizing radiation field and assumptions inherent in the Cloudy models. 

The most important result from this modeling analysis is that {\OVI} and {\NeVIII} do not occur in the cool ($T \aplt 10^4$~K) photoionized phase of the absorber. The photoionization model predictions for the {\OVI} column densities are $\aplt 3$~dex lower than the observed column density. Photoionization of {\NeVIII} requires extremely high ionization parameters of $\log~U \sim -0.8$ for even solar [Ne/H] abundances. The corresponding densities of $n_{\H} \sim 10^{-5}$~{\cc} lead to very large path lengths ($\apgt 1$~Mpc). Absorption over such large path lengths is unlikely to result in the kinematically simple line profiles seen for these high ions. Furthermore, the broadening induced by the Hubble flow on the absorption over such large distances will be at least twice the measured $b(\OVI) = 33$~{\kms}. The photoionization predictions for {\OVI} and {\NeVIII} are thus physically unrealistic. The offset in the velocity centroid of the {\OVI} line and its higher $b$-value in comparision with the low and intermediate ionization species are clear indications of the multi-phase gas composition. It is more likely that the {\OVI} and the {\NeVIII} are regulated by collisional ionzation in a {\it warm} plasma at temperatures between $\sim 10^5 - 10^6$~K. We discuss this possibility in the next section.

\subsection{Collisional Ionization of {\OVI} and {\NeVIII} in the Warm Gas}

We can extract the temperature for the warm phase  of the absorber from the measured value of $N(\NeVIII)/N(\OVI)$. In the simple collisional ionization models of \citet{gnat07} shown in Figure 5, we find that the $N(\NeVIII)/N(\OVI) \sim 1$ is satisfied for gas temperatures of $T \sim 5.3 \times 10^5$~K. This temperature estimate is independent of the column density of {\HI} in this phase, but assumes a solar ratio for (Ne/O) relative abundance. The temperature corresponds to a thermal line broadening of $b_\mathrm{t}(\OVI) \sim 23$~{\kms} which is consistent with the measured $b(\OVI) = 33~{\pm}~2$~{\kms}, and indicates roughly equal contributions from thermal and non-thermal line broadening. The $T \sim 5.2 \times 10^5$~{K} will be a lower limit if we account for the possibility that the {\OVI} can have some contribution from the photoionized gas as well. The intermediate ions like {\CIII}, {\NIII}, {\NIV}, {\OIII}, and {\OIV} have very low ionization fractions ($f \aplt 10^{-5}$) at such temperatures and will contribute negligibly to the total column density.

The {\HI} associated with the warm phase will be broad, dominated by its thermal $b$-value of $ \sim 93$~{\kms}. In other words, it will be a broad-{\Lya} absorber (BLA). The very low ionization fraction of $f(\HI) = N(\HI)/N(\H) = 5.85 \times 10^{-7}$ would result in the BLA absorption being shallow. This broad component would fall on top of the strong and narrow absorption from the photoionized gas. There is some hint for the presence of such a component in the wings of the {\Lya} line between velocities of [-110, -75]~{\kms} and [+75, +110]~{\kms}. The $S/N$ in this region is not adequate to clearly distinguish this feature. Nonetheless, as we show in figure 6, the observed {\Lya} profile is consistent with the presence of a broad component with $b(\HI) = 93$~{\kms} and $\log~N(\HI) \aplt 13.3$~dex.  This upper limit on the BLA column density corresponds to a lower limit of [X/H] $ \apgt -0.3$ on the true metallicity (see figure 5). The limiting BLA column density implies a total hydrogen column density of $\log~N(\H) \sim 19.5$~dex, a $\sim 2$ orders of magnitude higher baryon column density compared to the cool photoionized gas. From this analysis, we conclude a $\log~N(\HI) = 13.09~(-0.6, +0.3)$ with O and Ne abundances of [X/H] $\sim -0.12~ (-0.18, +0.12)$ in the warm collisionally ionized phase of the absorber. 

\subsection{A Constraint on the Broad-{\Lya} Absorption Tracing the Warm Gas}

The metallicity in this warm phase of the absorber is not well constrained since we do not have a direct measure on the associated {\HI}. If the warm gas is spatially coupled with the cooler photoionized gas, we can accept near-solar abundances for O and Ne as well. At solar metallicities and at $T \sim (2 - 7) \times 10^5$~K the metal cooling efficiencies can be high leading to recombination lags and shifts in ionization fractions of high ions from the collisional ionization equilibrium (CIE) predictions. However, as shown in Figure 5, the predictions from the non-equilibrium cooling models are similar to CIE. At solar metallicity, the single phase model would predict the measured $N(\NeVIII)$ and $N(\OVI)$ for $\log N(\HI) = 12.7$~dex. Alternatively, one can place an upper limit on the {\HI} column density, and a corresponding lower limit on the metallicity using the {\Lya}. 

We applied a formal fit to the {\Lya} by fixing the two narrow components of the {\HI} core (given in Table 3) and the velocity of the BLA at $v = 0$~{\kms} corresponding to the velocity centroid of the {\OVI} doublet lines. The fit yields a $b(\HI) = 86~{\pm}~6$~{\kms} and $\log~[N(\HI)] = 13.25~{\pm}~0.17$~dex. Considering the low detection significance of the BLA, the fitting procedure has underestimated the errors. The derived fit parameters are sensitive to the choice of the continuum and the assumptions on the component structure made on the basis of profile fitting. Regardless of this, it is interesting to note that the $b$ from the best-fit model is comparable to the value expected for $b(\HI)$ from gas at $T \sim 5.2 \times 10^5$~K. 

\section{Summary} 

We have added new insights into the gas phase properties of the $z = 0.32566$ {\NeVIII} absorber with the help of $HST$/COS high $S/N$ spectroscopic observations of 3C~263. The $FUSE$ detection of {\NeVIII} was reported by \citet{narayanan09}. The COS spectrum with coverage over the wavelength range of $1136 - 1796$~{\AA} shows lines from {\HI} ({\Lya} to {$\theta$}), {\OVI}, {\CIII}, {\NIII}, {\SiIII}, {\CII} at $z = 0.32566$, in addition to useful upper limits from non-detections of {\NII} and {\SiII}. This is supplemented by archival $FUSE$ observations of {\OIII}, {\OIV}, {\NIV} and {\NeVIII}. The main conclusion is that the {\NeVIII} combined with {\OVI} and possible broad {\HI} in this absorber are diagnostic of collisionally ionized gas with $T \sim 5.3 \times 10^5$~K. The other significant results are summarized as follows: 

\begin{enumerate}

\item The $z = 0.32566$ absorber is a multiphase mix of low ionization gas at $T \aplt 10^4$~K and {\it warm} high ionization gas at $T 5 \times 10^5$~K. Absorption in the low ionization gas shows at least two components at $v \sim -23$ {\kms} and $+12$~{\kms} in the the higher order Lyman lines and in the {\CIII}~977~{\AA} line. The COS spectrum also shows {\NIII}, {\SiIII}, and {\CII} associated with the positive velocity component. The $b$-values of these intermediate and low ions are $\aplt 10$~{\kms} implying low temperatures. The {\OVIdblt}~{\AA} lines are a factor of $\sim 4$ broader and are not kinematically aligned with the low, intermediate ions or the core of the {\HI} profile. 

\item The low and intermediate ions are consistent with an origin in gas photoionized by the extragalactic background radation. The bulk of the observed {\HI} is also traced by this photoionized medium. Simple photoionization models predict the measured low and intermediate ion column densities for a $\log~U \sim -2.2$ corresponding to a density of $n_{\H} \sim 5 \times 10^{-4}$~{\cc}, and a total hydrogen column density of $N(\H) \sim 2 \times 10^{18}$~{\cmsq}. The abundances in the photoionized phase are [C/H] $ = 0~{\pm}~0.04$~dex, [N/H] $= -0.3~{\pm}~0.1$, [O/H] $= -0.8~{\pm}~0.1$~dex, and [Si/H] $= -0.6~{\pm}~0.3$~dex.

\item The {\OVI} and {\NeVIII} favor an origin in collisionally ionized gas. The $N(\OVI) \sim N(\NeVIII)$ is predicted at $T = 5.2 \times 10^5$~K in CIE and non-equilibrium cooling models. The {\HI} absorption associated with this warm absorber is a BLA with $b(\HI) \sim 93$~{\kms} and comes from the trace neutral fraction ($f(\HI) = N(\HI)/N(H) = 5.85 \times 10^{-7}$) of hydrogen. The BLA is only marginally detected in the COS spectrum. From the observed {\NeVIII}, {\OVI} and the constraints set by the {\Lya} profile, we estimate for the warm gas phase a metallicity of [X/H] $\sim -0.12~(-0.18, +0.12)$, and a total hydrogen column density of $N(\H) = 3 \times 10^{19}$~{\cmsq}. The warm absorber contains a factor of $\sim 2$~dex more baryons than what is traced by the photoionized gas. 

\end{enumerate}

The ionization properties of the warm gas in this absorber are consistent with those predicted for the warm component of the WHIM, although the near-solar chemical abundances for carbon are higher than what is expected for the IGM. The absorber could be kinematically associated with halo gas. Without deep galaxy redshift measurements for the field surrounding 3C~263,  it will be difficult to draw firm conclusions about the actual physical site of the absorption. In Table 5, we have summarized the properties of the currently known population of {\OVI} - {\NeVIII} absorbers. Except for the sight line discussed in this paper, imaging data exists for all the other fields. In four out of the six remaining instances, relatively bright galaxies ($0.01 - 1~L^*$) were found proximate ($\sim 200$~kpc) to the absorbers. This is consistent with the correlation of {\OVI} absorbers with galaxies \citep[e.g.,][]{wakker09}, particularly their higher coveration fractions around galaxies that show evidence for star formation \citep{chen09, tumlinson11b}. 

We note that there have been few detections of {\NeVIII} - {\OVI} absorbers tracing $T \sim (0.5 - 1) \times 10^6$~K gas. One would expect an occasional imprint of the {\it hot} component of the WHIM in quasar spectra by way of absorbers with $N(\NeVIII) > N(\OVI)$ (see Figure 6). The apparent dearth of $T \sim 10^6$~K {\NeVIII} systems is puzzling, although the current statistics are small. A straightforward interpretation is that the {\NeVIII} absorbers in the current sample are having a physical origin different from the filamentary structures of the WHIM outside the virial boundaries of galaxies. The near-solar abundances are consistent with their origin in the extended regions around galaxies. Galaxy simulations find that the inclusion of feedback from star-formation and AGNs can heat circumgalactic gas to the warm temperatures of the WHIM \citep{cen06, tepper12}. In the simulations by \citet{tepper11}, majority of halo {\OVI} absorbers are tracing such high-metallicity regions enriched by supernova driven outflows where the gas has started to radiatively cool from the post-shock temperatures of $T \apgt 10^6$~K. 

The multiphase mix of cool and warm gas phases found in the {\NeVIII} - {\OVI} absorbers are also analogus to the highly ionized Milky Way high velocity clouds. The {\NeVIII} could have an origin similar to {\OVI} in the interface layers between the $T \aplt 10^4$~K high velocity gas and the $T \apgt 10^6$~K coronal ISM surrounding the galaxies \citep{sembach03, fox04, fox06}.  The properties of several extragalactic {\OVI} absorbers are found to be consistent with ionization in such conductive interfaces or mixing layers \citep{narayanan10a, savage10, savage11, tumlinson11a, tripp11}. The transition temperatures at the layers between the cold and hot phases would explain the narrow range in temperature probed by the current sample of {\NeVIII} absorbers (see Table 5). 

The high sensitivity spectra afforded by COS will possibly reveal many more {\NeVIII} - {\OVI} warm absorbers in the low redshift universe. Understanding where those absorbers reside with respect to galaxies along with a measurement on their chemical abundances will be important while predicting the origin of the warm gas. High metallicties and proximity to galaxies would favor an origin for the {\NeVIII} and {\OVI} in virialized halos rather than the canonical WHIM distant from galaxies.  

{\small This research is supported by the NASA Cosmic Origins Spectrograph program through a sub-contract to the University of Wisconsin-Madison from the University of Colorado, Boulder and NNX08-AC14G to the University of Colorado, Boulder. B.P.W. acknowledges support from NASA grant NNX-07AH426. The authors thank Jim Green \& team for delivering the Cosmic Origins Spectrograph and the STS-125 team for its installation on the HST. AN is indebted to Srianand Raghunathan and Sowgat Muzahid for valuable discussions. We are grateful to Gary Ferland and collaborators for developing the Cloudy photoionization code. We also thank Orly Gnat for making the computational data on radiatively cooling models public. This research has made use of the NASA/IPAC Extragalactic Database (NED) which is operated by the Jet Propulsion Laboratory, California Institute of Technology, under contract with the National Aeronautics and Space Administration.}

\pagebreak


\clearpage
\begin{deluxetable}{lcccccc}
\tablecaption{\textsc{COS Observations of 3C~263}}
\tablehead{
\colhead{HST ID} &
\colhead{Obs.Date} &
\colhead{Grating} &
\colhead{FP-POS} &
\colhead{$\lambda_c$} &
\colhead{$\Delta \lambda$} &
\colhead{$t_{exp}$} \\
\colhead{ } &
\colhead{(yyyy:mm:dd)} &
\colhead{ } &
\colhead{ } &
\colhead{(\AA)} &
\colhead{(\AA)} &
\colhead{(sec)} 
}
\startdata
LB6810010	&	2010-01-01		&	G130M	&	3	&	1291	&	$1135 - 1427$	&	3840\\
LB6810020	&	2010-01-01		&	G130M	&	3	&	1300	&	$1145 - 1437$	&	3840\\
LB6810030	& 2010-01-01		&	G130M	&	3	&	1309	&	$1155 - 1446$	&	3840\\
LB6810040	&	2010-01-01		&	G130M	&	3	&	1318	&	$1165 - 1456$	&	3840\\
LB6811020	&	2010-02-22		&	G160M	&	3	&	1589	&	$1165 - 1459$	&	4500\\
LB6811030	&	2010-02-22		&	G160M	&	3	&	1600	&	$1147 - 1440$	&	4500 \\
LB6811040 & 2010-02-22    & G160M & 3 & 1611  & $1425 - 1782$ & 4500\\
LB6811050 & 2010-02-22    & G160M & 3 & 1623  & $1437 - 1795$ & 4500
\enddata
\tablecomments{ Column (1) is the HST ID for the data set, column (2) gives the date of observation, column (3) lists the choice of grating, columns (4) and (5) are the the FP position and central wavelength of the grating, column (6) gives the wavelength range sampled under each setting and column (7) lists the duration of the exposure. All observations were done as part of the COS-GTO program ID 11541 (PI Green).}  
\label{tab:tab1}
\end{deluxetable}

\begin{deluxetable}{lrrrr}
\tablewidth{0pt}
\tablecaption{\textsc{AOD Measurements of Lines in COS Spectra}}
\tablehead{
\colhead{Line} &
\colhead{$W_r$} &
\colhead{$b_a$} &
\colhead{log~$[N_a~(\cmsq)]$} &
\colhead{$[-v, +v]$}  \\
\colhead{ } &
\colhead{(m\AA)} &
\colhead{(\kms)} &
\colhead{dex} &
\colhead{(\kms)}
}
\startdata
{\HI}~1215      & $> 453$             &   $51~{\pm}~4$    &  $> 14.3$ & [-150,150] \\
{\HI}~1025      & $> 314$             &   $42~{\pm}~4$    &  $> 15.0$ & [-150,100] \\
{\HI}~972       & $> 256$             &   $46~{\pm}~5$    &  $> 15.3$ & [-150,100] \\
{\HI}~938 		  &	$110 ~{\pm}~ 8$     & 	$37 ~{\pm}~ 4$ 	&  $15.36 ~{\pm}~ 0.04$ 	& 	[-75,75] 	\\
{\HI}~930       & $73~{\pm}~6$        &   $37~{\pm}~4$    &  $15.36 ~{\pm}~ 0.03$   & [-75,75] \\
\\
{\OVI}~1031 		& $91 ~{\pm}~ 9$ 	    & 	$37 ~{\pm}~ 4$ 	& 	$13.94 ~{\pm}~ 0.05$ 	& 	[-75,75] 	\\
{\OVI}~1037 		& $52 ~{\pm}~ 6$ 		  & 	$42 ~{\pm}~ 7$ 	& 	$13.96 ~{\pm}~ 0.04$  & 	[-75,75] 	\\
\\
{\CIII}~977 		&	$179 ~{\pm}~ 7$ 	  & 	$38 ~{\pm}~ 4$ 	& 	$13.60 ~{\pm}~ 0.04$ 	& 	[-75,75] 	\\
{\CIII}~977 		& $78 ~{\pm}~ 9$ 	    & 	$18 ~{\pm}~ 7$	& 	$13.27 ~{\pm}~ 0.04$ 	& 	[-75,-10] \\
{\CIII}~977 		& $95 ~{\pm}~ 10$ 	  & 	$26 ~{\pm}~ 4$ 	& 	$13.35 ~{\pm}~ 0.04$  &	  [-10,75] 	\\
\\
{\NIII}~989 		& $37 ~{\pm}~ 10$ 	  & 	$27 ~{\pm}~ 4$ 	& 	$13.61 ~{\pm}~ 0.05$	& 	[-10,75] \\
{\NIII}~989 		& $< 48$ 	            & 	$ ... $ 	      & 	$< 13.1$  	          &	  [-75,-10] \\
\\
{\SiIII}~1206   & $74 ~{\pm}~ 7$      & $30 ~{\pm}~ 3$  &   $12.64 ~{\pm}~ 0.09$  & [-40, 40] \\
{\SiIII}~1206 	& $49 ~{\pm}~ 7$      & $17 ~{\pm}~ 3$ 	& 	$12.46 ~{\pm}~ 0.05$ 	& 	[-10,40] \\
{\SiIII}~1206 	& $27 ~{\pm}~ 7$ 		  & $17 ~{\pm}~ 3$  & 	$12.17 ~{\pm}~ 0.08$ 	& 	[-40,-10] \\
\\
{\CII}~903a 		& $24 ~{\pm}~ 6$ 	    & 	$14 ~{\pm}~ 3$ 	& 	$13.07 ~{\pm}~ 0.05$ 	& 	[-15,30] \\
{\CII}~903b 		& $17 ~{\pm}~ 5$ 	    & 	$15 ~{\pm}~ 2$ 	& 	$13.17 ~{\pm}~ 0.07$ 	& 	[-15,30] \\
{\CII}~1036 		& $16 ~{\pm}~ 6$ 	    & 	$13 ~{\pm}~ 3$ 	&   $13.18 ~{\pm}~ 0.07$  &	  [-15,30] \\
{\CII}~1036 		&  	$< 15$ 		        & 	 $  ...  $ 		  &  	$< 13.1$			        & 	[-75,-10] \\
\\
{\NII}~1083 		&  	$< 12$		        &		 $...$ 		      &  	  $< 13.0$			      & 	[-75,75] \\
\\
{\SiII}~1193 		&  	$< 28$		        & 	 $...$ 		      &  	  $< 12.8$			      & 	[-75,75] 
\enddata
\tablecomments{The AOD integrations were performed over the velocity range over which absorption is seen. The {\Lya}, {\Lyb} and {\Lyg} have saturated line cores and the measurements are therefore lower limits. The {\CIII}~977~{\AA} and {\SiIII}~1206~{\AA} lines have a two component profile. The AOD measurements were done over the full absorption as well as the positive and negative velocity components. {\NIII}, and {\CII} show absorption only in the positive velocity component. We have listed $3\sigma$ upper limits for the negative velocity component. The {\CII} $\lambda = 903.9616$~{\AA} ($f_{osc} = 0.336$) transition is labelled as {\CII}~903a and the $\lambda = 903.6235$~{\AA} ($f_{osc} = 0.168$) transition as {\CII}~903b. {\NII} and {\SiII} are not detected at the $> 3\sigma$ level.}
\label{tab:tab2}
\end{deluxetable}

\clearpage
\begin{deluxetable}{lrrrrr}
\tablewidth{0pt}
\tablecaption{\textsc{Profile Fit Measurements of Lines in COS Spectra}}
\tablehead{
\colhead{Line} &
\colhead{$v$} &
\colhead{log~$[N~(\cmsq)]$} &
\colhead{$b$} &
\colhead{$\chi^2_\nu$} \\
\colhead{ } &
\colhead{(\kms)} &
\colhead{dex} &
\colhead{(\kms)} &
\colhead{ }
}
\startdata
${\Lya}, {\beta}, {\gamma}, {\epsilon}, {\zeta}, {\eta}, {\theta}$  &  $-28~{\pm}~4~{\pm}~8$  &  $14.99~{\pm}~0.04$ & $18~{\pm}~3$  &  1.14\\
                                                                    &  $7~{\pm}~3~{\pm}~8$    &  $15.20~{\pm}~0.04$ & $17~{\pm}~3$  &   \\
\\
${\Lya}, {\beta}, {\gamma}, {\epsilon}, {\zeta}, {\eta}, {\theta}$  &  $-23$  &  $15.13~{\pm}~0.05$ & $21~{\pm}~2$ & 1.17 \\
                                                                    &  $12$   &  $15.09~{\pm}~0.05$ & $15~{\pm}~3$ & \\
\\
{\CIII}~977       &   $-23~{\pm}~2~{\pm}~8$   &   $13.45~{\pm}~0.05$  &   $12~{\pm}~2$  & 0.56 \\
                  &   $12~{\pm}~2~{\pm}~8$    &   $13.67~{\pm}~0.09$  &   $9~{\pm}~3$  & \\
\\
{\CII}~903a       &   $13~{\pm}~2~{\pm}~8$    &   $13.23~{\pm}~0.05$  &   $10~{\pm}~3$   & 0.40 \\
\\
{\CII}~903b       &   $13~{\pm}~2~{\pm}~8$    &   $13.30~{\pm}~0.08$  &   $6~{\pm}~3$   & 0.30 \\
\\
{\CII}~1036       &   $13~{\pm}~1~{\pm}~8$    &   $13.24~{\pm}~0.07$  &   $6~{\pm}~3$   & 0.63 \\
\\
{\CII}~903a,903b,1036 &   $13~{\pm}~2~{\pm}~8$    &   $13.26~{\pm}~0.03$  &   $8~{\pm}~2$   & 0.53 \\
\\
{\NIII}~989       &   $12~{\pm}~1~{\pm}~8$    &   $13.75~{\pm}~0.04$  &   $8~{\pm}~2$   & 0.52 \\
\\
{\OVI}~1032/1038  &   $1~{\pm}~1~{\pm}~8$     &   $13.98~{\pm}~0.05$  &   $33~{\pm}~2$  & 0.53 \\
\\
{\SiIII}~1206     &   $16~{\pm}~2~{\pm}~8$    &   $12.76~{\pm}~0.29$  &   $5~{\pm}~3$   & 0.99 \\
                  &   $-20~{\pm}~4~{\pm}~8$   &   $12.18~{\pm}~0.10$  &   $12~{\pm}~3$  &  \\
\enddata
\tablecomments{The velocity uncertainty includes the statistical errors computed by the fitting routine and a ${\pm}~8$~{\kms} velocity calibration error in the COS spectrum. The {\HI} parameters were obtained from a simultaneous free-fit to the various uncontaminated Lyman series lines. The reduced-$\chi^2$ of the fit is affected by the wavelength calibrations errors in the spectrum. A simultaneous fit on {\HI} was also done by locking the velocity centroids of the components to the velocities in the {\CIII}~977~{\AA} absorption. The fit parameters from this and the free-fit are within $2\sigma$ of each other. The {\CII} $\lambda = 903.9616$~{\AA} ($f_{osc} = 0.336$) transition is labelled as {\CII}~903a and the $\lambda = 903.6235$~{\AA} ($f_{osc} = 0.168$) as {\CII}~903b. Simultaneous fits were done for the {\CII} multiplet lines and the {\OVIdblt}~{\AA} doublet lines.} 
\label{tab:tab3}
\end{deluxetable}

\begin{deluxetable}{lrrrr}
\tablewidth{0pt}
\small
\tablecaption{\textsc{AOD Measurements of Lines in $FUSE$ Spectra}}
\tablehead{
\colhead{Line} &
\colhead{$W_r$} &
\colhead{$b_a$} &
\colhead{log~$[N_a~(\cmsq)]$} &
\colhead{$[-v, +v]$} \\
\colhead{ } &
\colhead{(m\AA)} &
\colhead{(\kms)} &
\colhead{dex} &
\colhead{(\kms)} 
}
\startdata
{\OII}~834		&  $< 47$		      &	  $...$	      &   $< 13.8$		      & [-75,75] \\
\\
{\OIII}~833		&  $42~{\pm}~9$	  & $23~{\pm}~6$	&	$13.89~{\pm}~0.07$	&	[-75,-10] \\
{\OIII}~833		&  $51~{\pm}~8$	  & $22~{\pm}~11$	&	$14.00~{\pm}~0.06$	&	[-10,75] \\
\\
{\OIV}~788		&  $41~{\pm}~7$	  & $25~{\pm}~8$	& $13.93~{\pm}~0.06$	&	[-75,-10] \\
{\OIV}~788		&  $34~{\pm}~8$	  & $17~{\pm}~9$	& $13.85~{\pm}~0.07$	&	[-10,75] \\
\\
{\NIV}~765		&  $45~{\pm}~10$	& $27~{\pm}~4$	&	$13.23~{\pm}~0.08$	& [-75,-10]	\\
{\NIV}~765		&  $30~{\pm}~8$	  & $31~{\pm}~4$	&	$13.10~{\pm}~0.08$	& [-10,75]	\\
\\
{\NeVIII}~770	&  $34~{\pm}~11$	& $27~{\pm}~7$  & $13.87~{\pm}~0.08$  &  [-75,-10] \\
{\NeVIII}~770	&  $20~{\pm}~10$	& $37~{\pm}~6$  & $13.40~{\pm}~0.14$  &  [-10,75] \\
{\NeVIII}~770 &  $52~{\pm}~11$  & $52~{\pm}~7$  & $13.99~{\pm}~0.11$  &  [-75,75]
\enddata
\tablecomments{The $FUSE$ data has lower $S/N$ compared to COS and therefore the component sub-structure seen in the COS absorption lines for this system are not evident. For the purpose of modeling we have split the integration range of $N_a(v)$ into two regions corresponding to the velocities of the two components in {\HI} and {\CIII} seen in the COS spectrum.}
\label{tab:tab4}
\end{deluxetable}

\clearpage
\begin{deluxetable}{lrcccccrc}
\tablewidth{0pt}
\small
\tablecaption{\textsc{Properties of Known {\OVI}-{\NeVIII} Intervening Absorbers}}
\tablehead{
\colhead{QSO} &
\colhead{$z$} &
\colhead{$\log~N(\OVI)$} &
\colhead{$\log~N(\NeVIII)$} &
\colhead{$\log~T$} &
\colhead{$\log~N(\H)$} &
\colhead{[X/H]} &
\colhead{Assoc. Galaxy} &
\colhead{Note}
}
\startdata
HE~0226-4110  & 0.2070 & $14.37~{\pm}~0.03$  & $13.89~{\pm}~0.11$  & 5.70  & $\sim 20.1$  & $\sim -0.9$ & 109 kpc, 0.25 $L^*$ & 1 \\
PKS~0405-123  & 0.4951 & $14.39~{\pm}~0.01$  & $13.96~{\pm}~0.06$  & 5.72  & $\sim 19.7$ & $\sim -0.6$  & 110 kpc, $0.08~L^*$ & 2 \\
PG~1148+549 & 0.6838  & $14.47~{\pm}~0.03$  & $13.98~{\pm}~0.09$  & 5.68  & $\sim 19.8$ & $> -0.5$ & ... & 3 \\
            & 0.7015  & $14.29~{\pm}~0.04$  & $13.75~{\pm}~0.07$  & 5.69  & $\sim 19.2$ & $\apgt 0$ & ... & 3 \\ 
            & 0.7248  & $13.84~{\pm}~0.10$  & $13.70~{\pm}~0.12$  & 5.72  & $\sim 18.8$ & $\apgt 0$ & 217 kpc, $L^*$ & 3 \\
3C~263  & 0.3257  & $13.98~{\pm}~0.05$  & $13.99~{\pm}~0.11$  & 5.72  & $\sim 19.3$ & $\apgt -0.1$ & ... & 4  \\
PG~1206+459   & $\sim 0.927$ &  $ ... $  & $\sim 14.9$ & $\sim 5.6$  & $\sim 19.8$ & $\apgt 0$ & 68 kpc, $1.8~L^*$ & 5 
\enddata
\tablecomments{(1) From \citet{savage11}; the abundance listed is for oxygen. (2) From \citet{narayanan11}; the abundance listed is for neon; the total hydrogen column density was found to be in the range $19 - 20$~dex. (3) The $z = 0.6838$ and $z = 0.7015$ absorbers in \citet{meiring12} have no associated galaxies detected down to $m_\mathrm{U} < 27$. (4) This paper; there is no galaxy information available for this sightline. (5) The absorption has several components kinematically spread over $\sim 1450$~{\kms}. The {\NeVIII} column density listed is the total column density from all the components as given in \citet{tripp11}. The temperature, metallicity and information on the galaxy are based on \citet{tripp11}. The {\OVI} lines are covered only in the lower resolution (FWHM $\sim 230$~{\kms}) $HST$/FOS spectrum and reported in \citet{ding03}. The absorption is very strong [$W_r(\OVI~1032) \sim 0.5$~{\AA}], but there are no column density measurements because of various line contamination issues.}
\label{tab:tab5}
\end{deluxetable}


\clearpage
\begin{figure*}
\begin{center}
\includegraphics[scale=0.9]{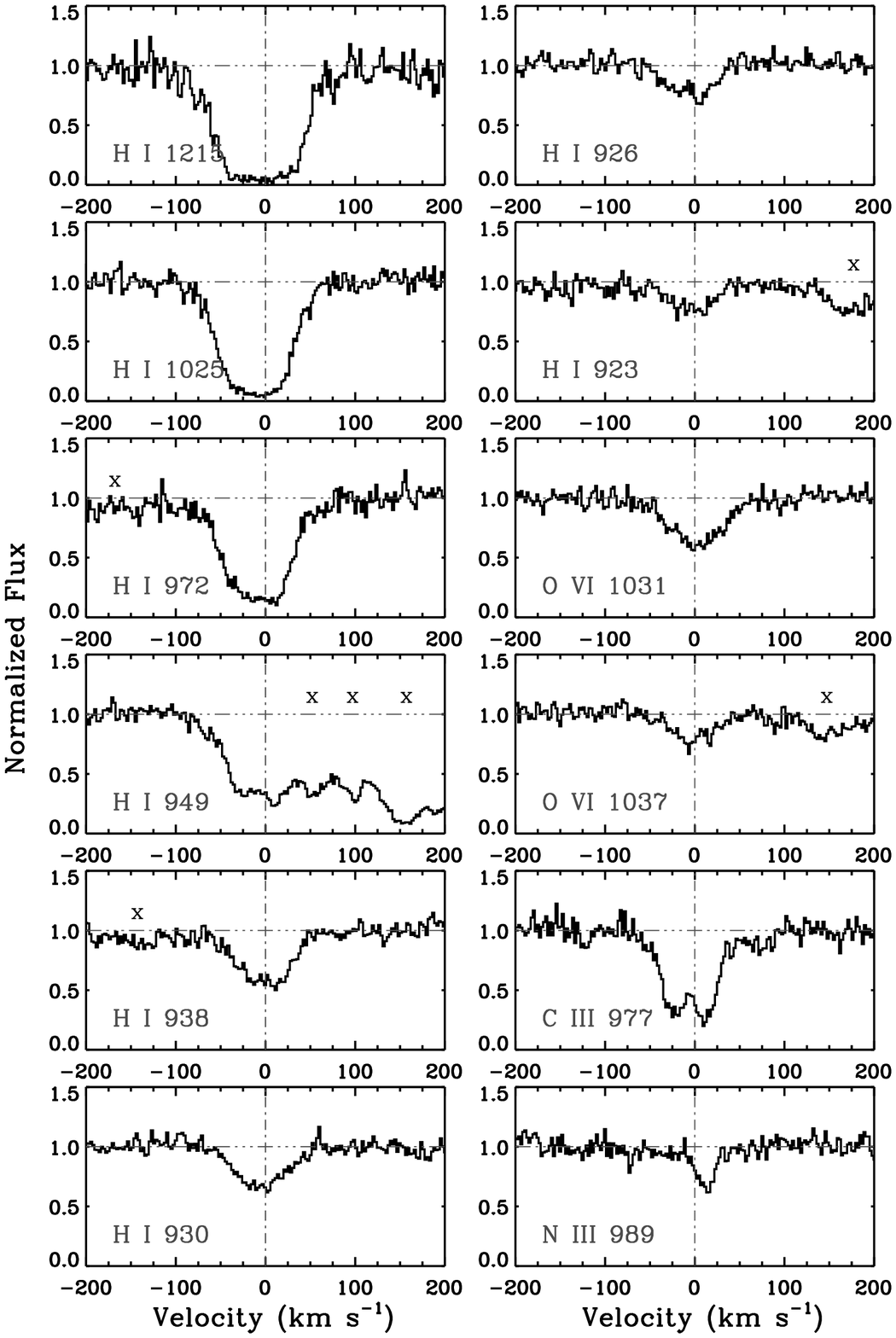}
\end{center}
\protect
\vspace{0.3in}
{\large Fig.~1a~--~Continuum normalized COS and $FUSE$ spectra of $3C~263$ where $v = 0$~{\kms} corresponds to $z = 0.32566$. Features which are not part of the absorption system are marked "x" in the corresponding panels. The AOD and profile fit line measurements are listed in Tables 2 \& 3. The Ly-$\delta$ ({\HI}~$950$) line is contaminated by Galactic {\SiII}~1260 and possibly also by {\SII}~1260. The {\CII} lines at $\lambda = 903.9616$~{\AA} and $903.6235$~{\AA} are labelled as {\CII}~903a and {\CII}~903b respectively.The {\NII}~1084 and {\SiII}~1193 transitions covered by COS and {\OII}~834 transition covered by $FUSE$ are not detected at $\geq 3\sigma$ significance.}
\label{fig:1}
\end{figure*}

\clearpage
\begin{figure*}
\begin{center}
\includegraphics[scale=0.9]{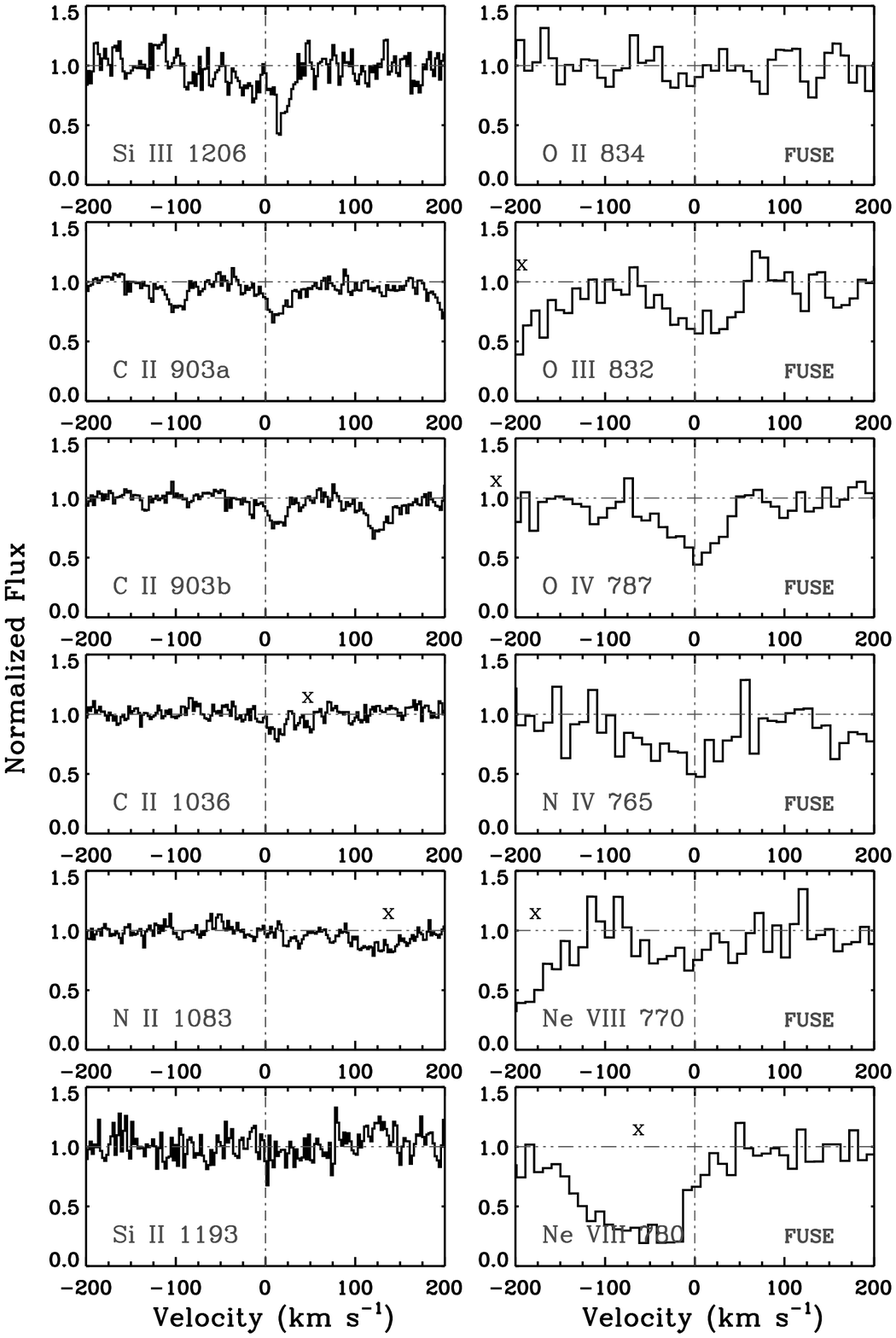}
\end{center}
\protect
{\large Fig.~1b~--~Continuation of Figure 1a.}
\label{fig:1}
\end{figure*}

\setcounter{figure}{1}
\clearpage
\begin{figure*}
\begin{center}
\includegraphics[scale=0.9]{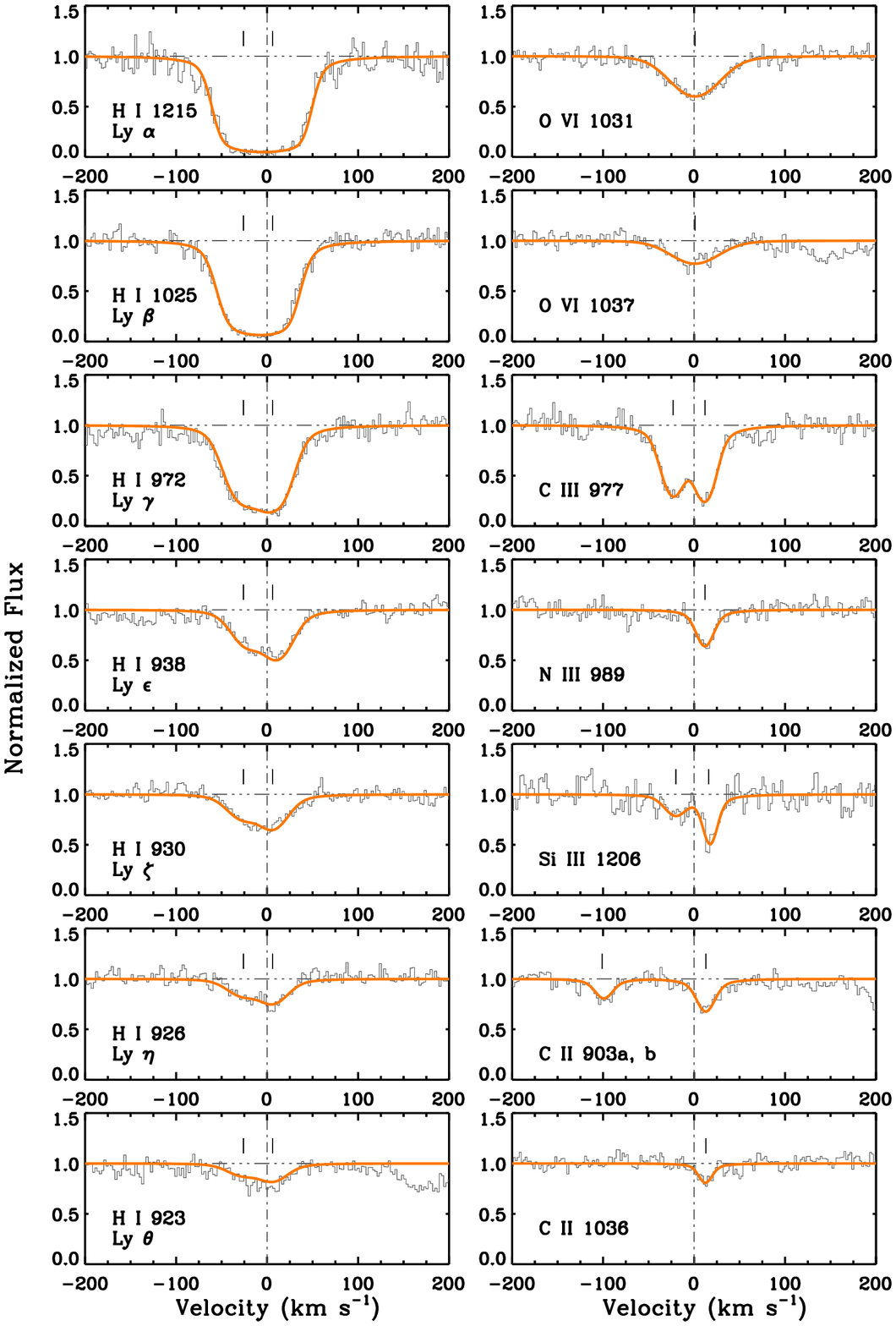}
\end{center}
\protect
\caption{\large Continuum normalized COS spectrum of 3C~$263$ of the $v = 0.32566$ absorber with Voigt profile models superimposed on the different lines. The fit parameters for the lines are given in Table 3. The vertical tick marks above each line represents the velocity centroid of the absorbing component. In the higher $S/N$ COS spectrum, the {\HI} lines and {\CIII} clearly show a two component structure. Coincident with the positive velocity component are also seen absorption from {\NIII}, {\SiIII}, and {\CII}. The negative velocity component of {\SiIII} is a weak feature. The {\OVIdblt} lines are offset in velocity compared to the low ions. The {\OVI} profiles are consistent with absorption from a single component which is $\sim 4$ broader in comparison with the low ions.}
\label{fig:2}
\end{figure*}

\clearpage
\begin{figure*}
\begin{center}
\vspace{-2.5in}
\includegraphics[scale=0.7]{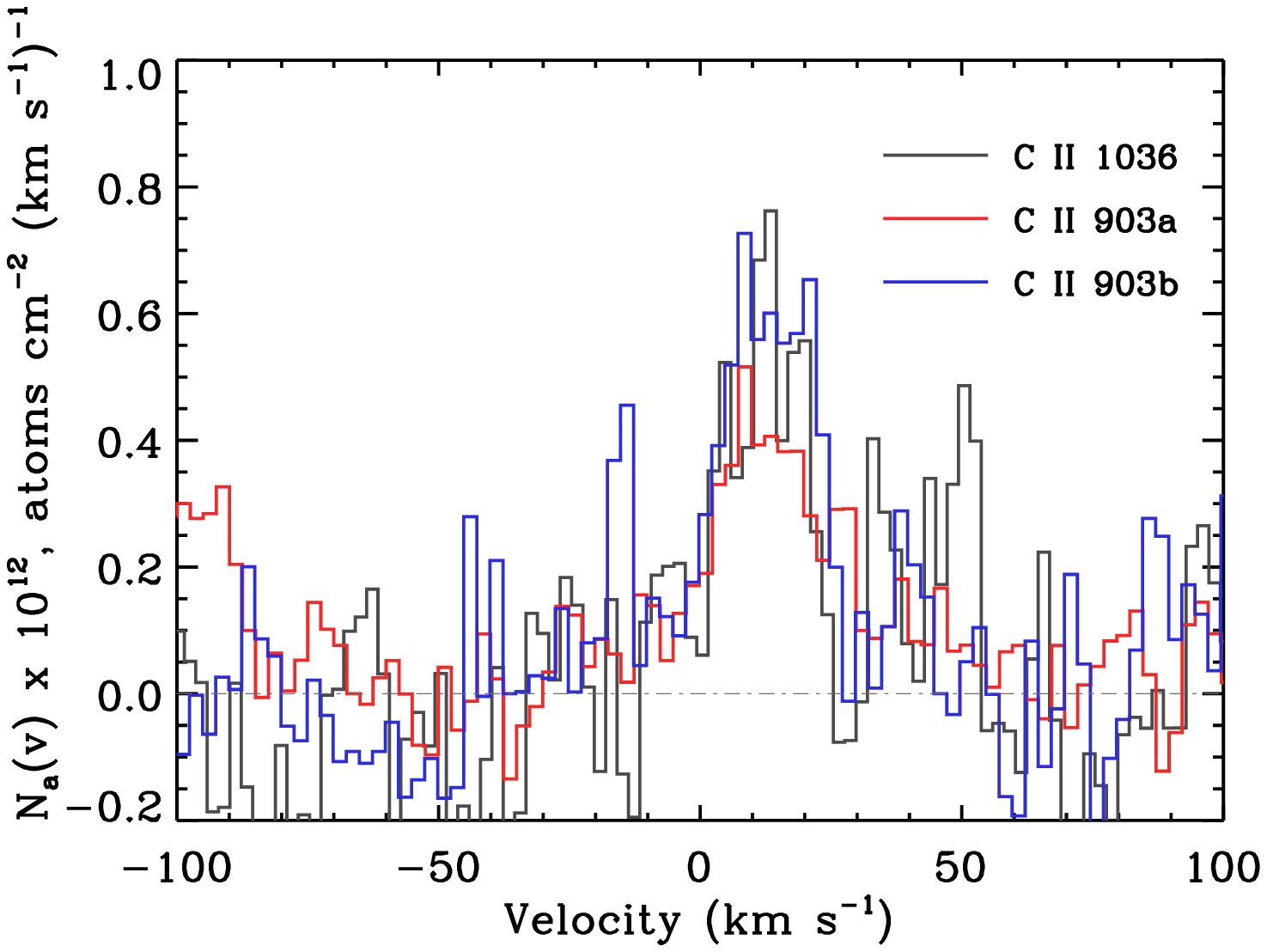} \\
\vspace{-4.0in}
\includegraphics[scale=0.7]{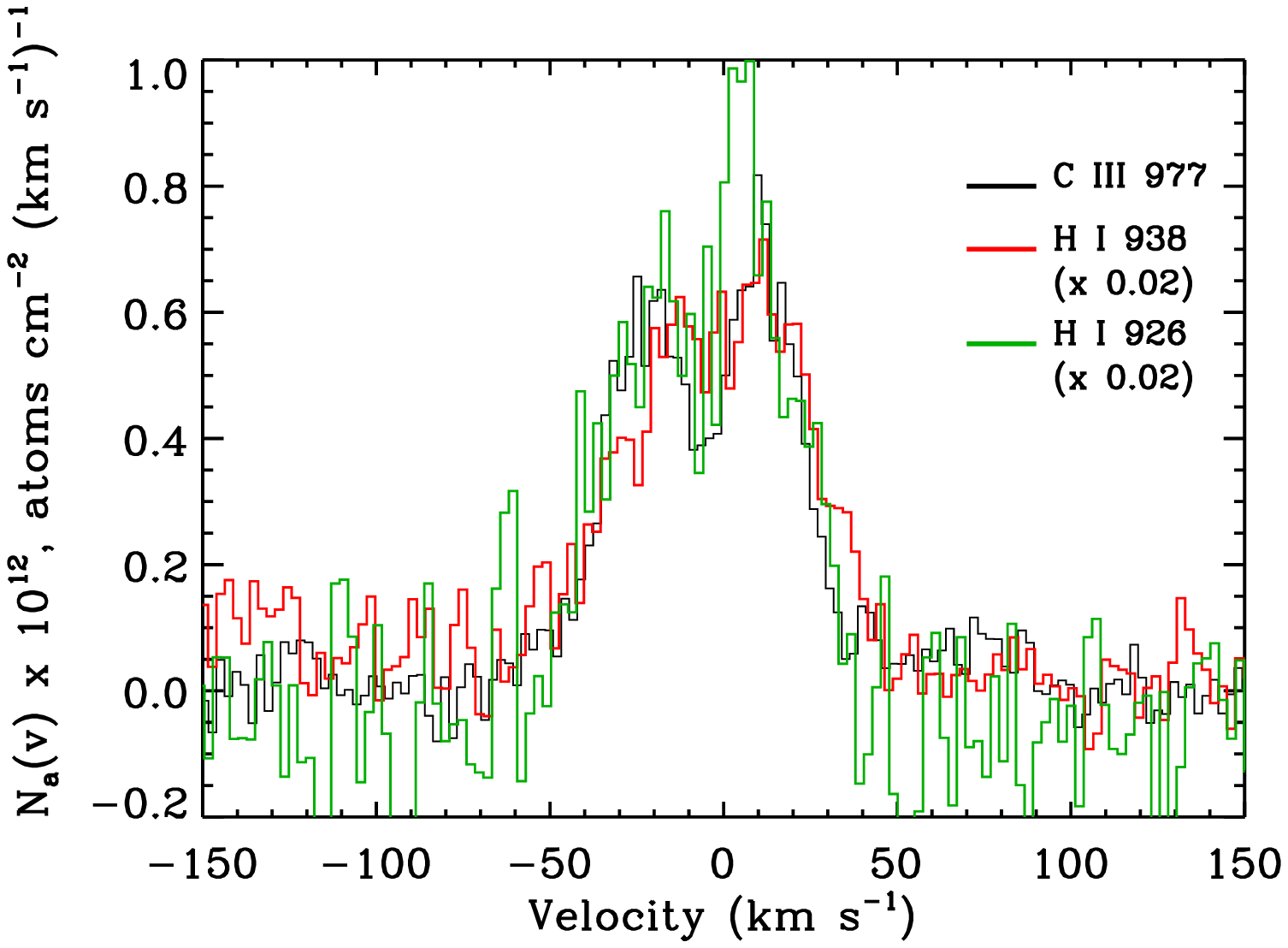}\\
\vspace{-4.0in}
\includegraphics[scale=0.7]{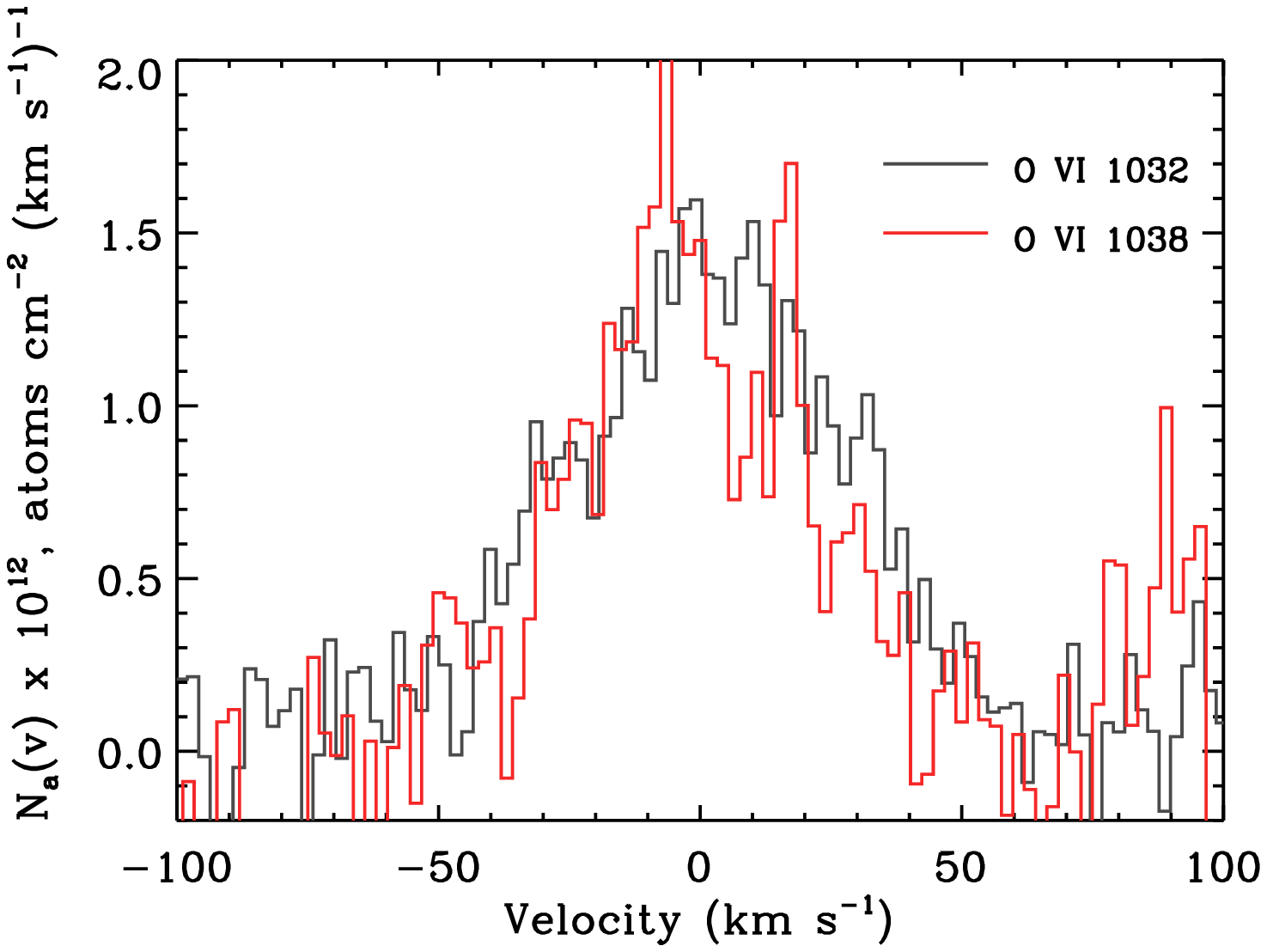}
\end{center}
\protect
\vspace{-1.7in}
\caption{The {\it top panel} shows the apparent column density comparison between {\CII}~904a ($\lambda = 903.9616$~{\AA}, $f_{osc} = 0.336$), {\CII}~904b ($\lambda = 903.6235$~{\AA}, $f_{osc} = 0.168$) and {\CII}~1036 ($\lambda = 1036.3367$~{\AA}, $f_{osc} = 0.1231$) transitions. The $N_a(v)$ comparison suggests that the lines are unresolved at the FWHM $\sim 17$~{\kms} resolution of COS and are narrower than the observed line widths. The $N_a(\CII)$ when corrected for this instrumental blurring is consistent with the profile fit value given in Table 3. In the {\it middle panel} we see the kinematic coincidence between the two component absorption in {\CIII} and {\HI}. The component structure in {\HI} is evident only in the unsaturated higher order Lyman lines. The {\CII} line aligns with the positive velocity component in {\CIII} and {\HI}. The similar $N_a(v)$ profiles of {\OVIdblt} lines shown in the {\it bottom panel} indicates the absence of contamination or unresolved saturation. The {\OVI} lines are kinematically broader and different compared to {\CIII} or {\CII}.}
\label{fig:3}
\end{figure*}

\clearpage
\begin{sidewaysfigure}
\begin{center}
\vspace{1in}
\includegraphics[scale=0.8,angle=90]{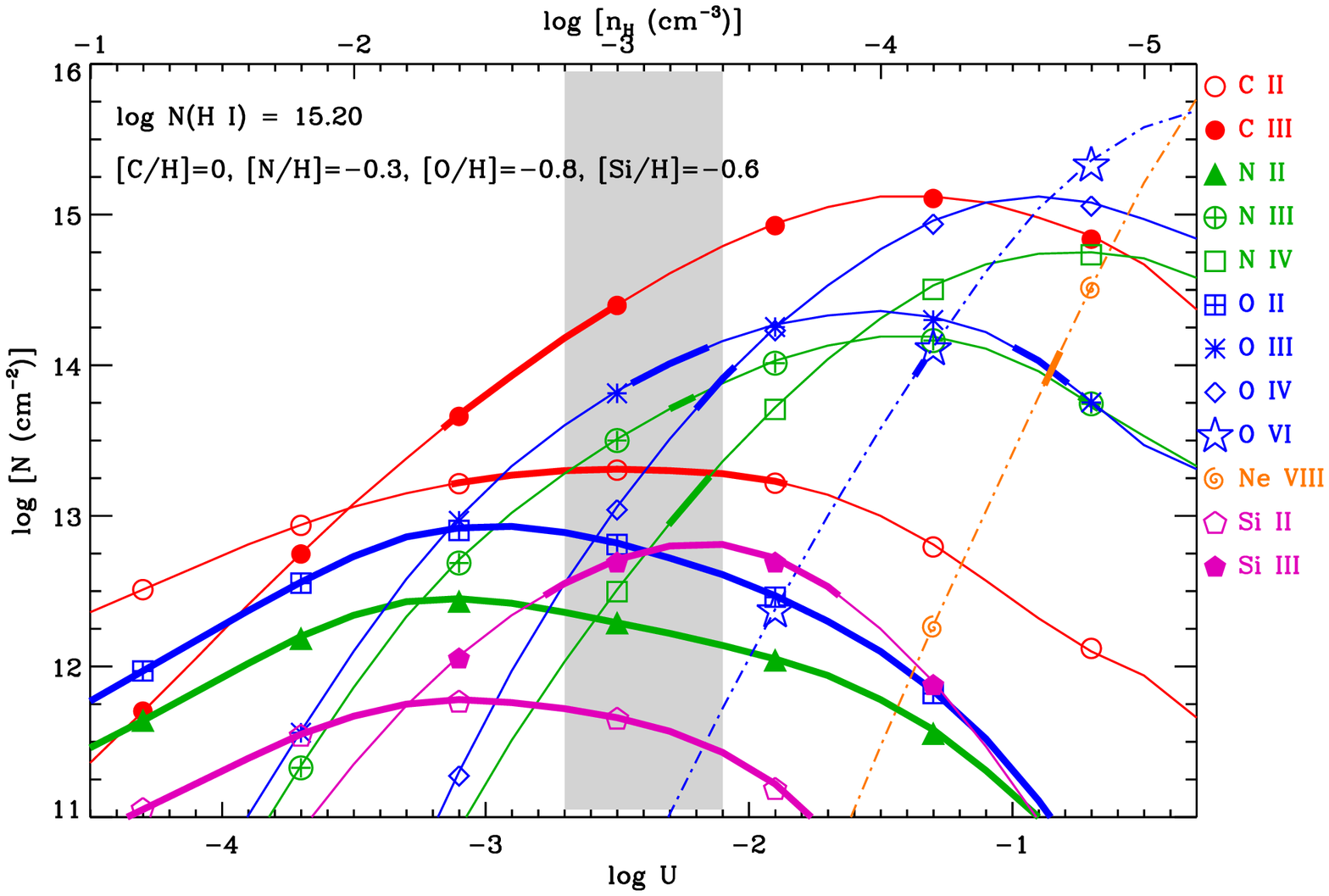}
\end{center}
\protect
\caption{\large Photoionization model predicted column densities for the positive velocity component of the $z = 0.32566$ absorber detected in {\HI}, {\CIII}, {\NIII}, {\SiIII}, {\CII} by COS and in {\OIII}, {\OIV} and {\NIV} by $FUSE$. The bottom X-axis is the ionization parameter $\log~U$, and the top X-axis is the total hydrogen number density $\log~[n_{\H}~\cc]$. The extragalactic ionizing background is from the Haardt \& Madau (2001) model for $z = 0.32566$. The background includes UV photons from quasars and young star-forming galaxies. The acceptable range of column densities for each ion are highlighted in the photoionization curves using thick lines. {\OII}, {\NII} and {\SiII} are non-detections at the $\geq 3\sigma$ significance level and the corresponding column density measuremenets are upper limits. The models are constructed for a $\log~N(\HI) = 15.20$~dex obtained from the profile fitting of {\HI} lines. The shaded box represents the range of $\log~U$ within which the values for photoionization models are a good match to the observed column densities of the low ions. To produce the observed column densities of the high ions ({\OVI}, {\NeVIII}) from photoionization requires physically unrealistic scenarios. The results are discussed in detail in Sec 5.} 
\label{fig:4}
\end{sidewaysfigure}

\clearpage
\begin{sidewaysfigure}
\begin{center}
\includegraphics[scale=0.9,angle=90]{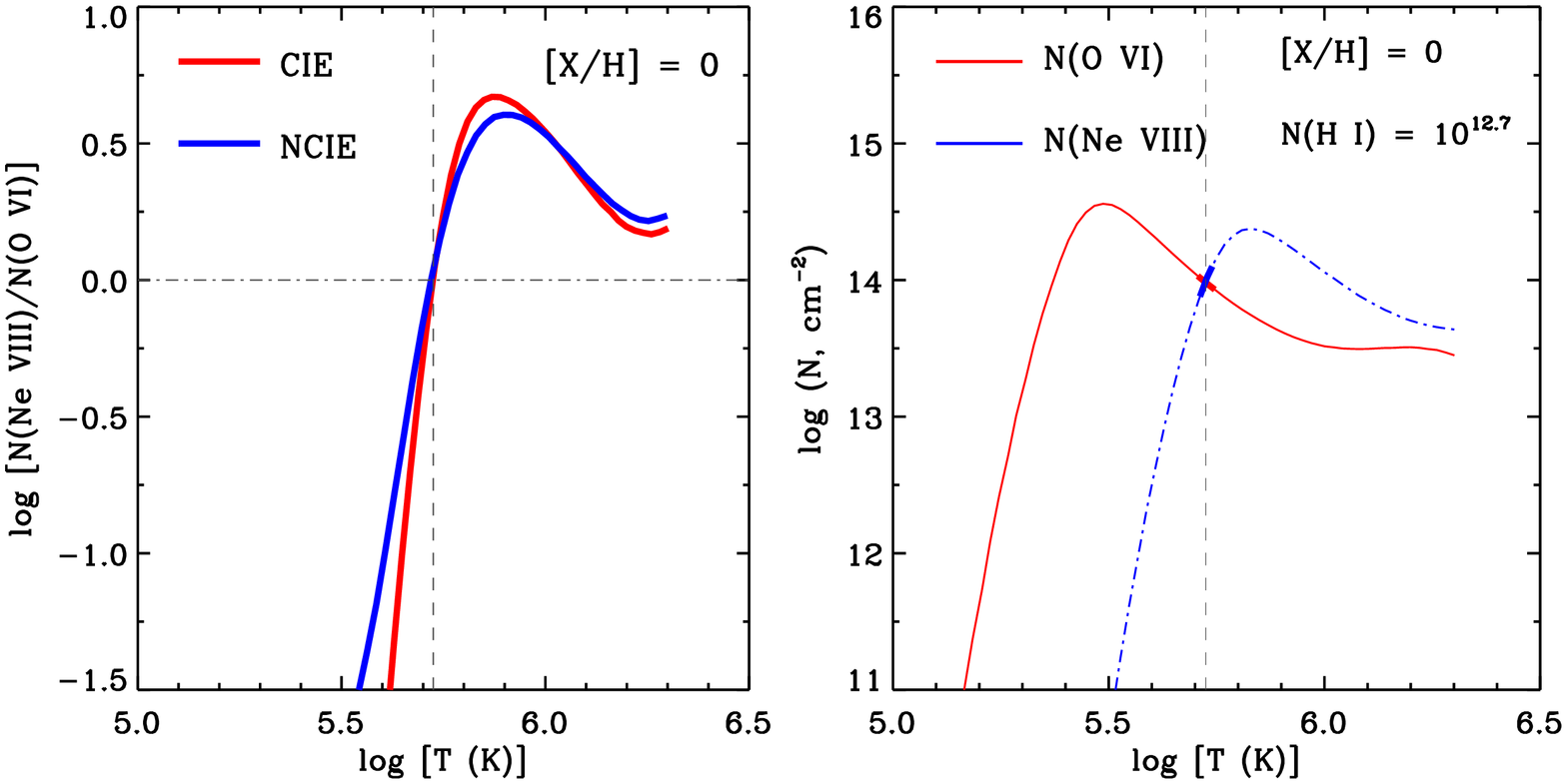}
\end{center}
\protect
\caption{\large {\it Left Panel :} Collisional ionization equilibrium ({\it red curve}) and non-equilibrium ({\it blue curve}) cooling model predictions for the column density ratio between {\NeVIII} and {\OVI} for a range of plasma temperatures, based on Gnat \& Sternberg (2007). The horizontal {\it dash-dot} line is the measured $N(\NeVIII)/N(\OVI)$ and the vertical line is the temperature at which the model predictions match this ratio. The predictions from CIE and non-CIE models are identical for this temperature. {\it Right Panel :} Column densities for {\NeVIII} and {\OVI} as a function of temperature from CIE models. For the models to reproduce the $N(\NeVIII)$ and $N(\OVI)$ at $\log$ T $= 5.72$ (T $ = 5.2 \times 10^5$~K) at solar metallicity, the column density of {\HI} has to be $\log~N(\HI) = 12.7$~dex. This can be considered as a lower limit on the {\HI} column density, assuming that this warm phase cannot have supersolar metallicities.}
\label{fig:5}
\end{sidewaysfigure}

\clearpage
\begin{figure*}
\begin{center}
\includegraphics[scale=0.9]{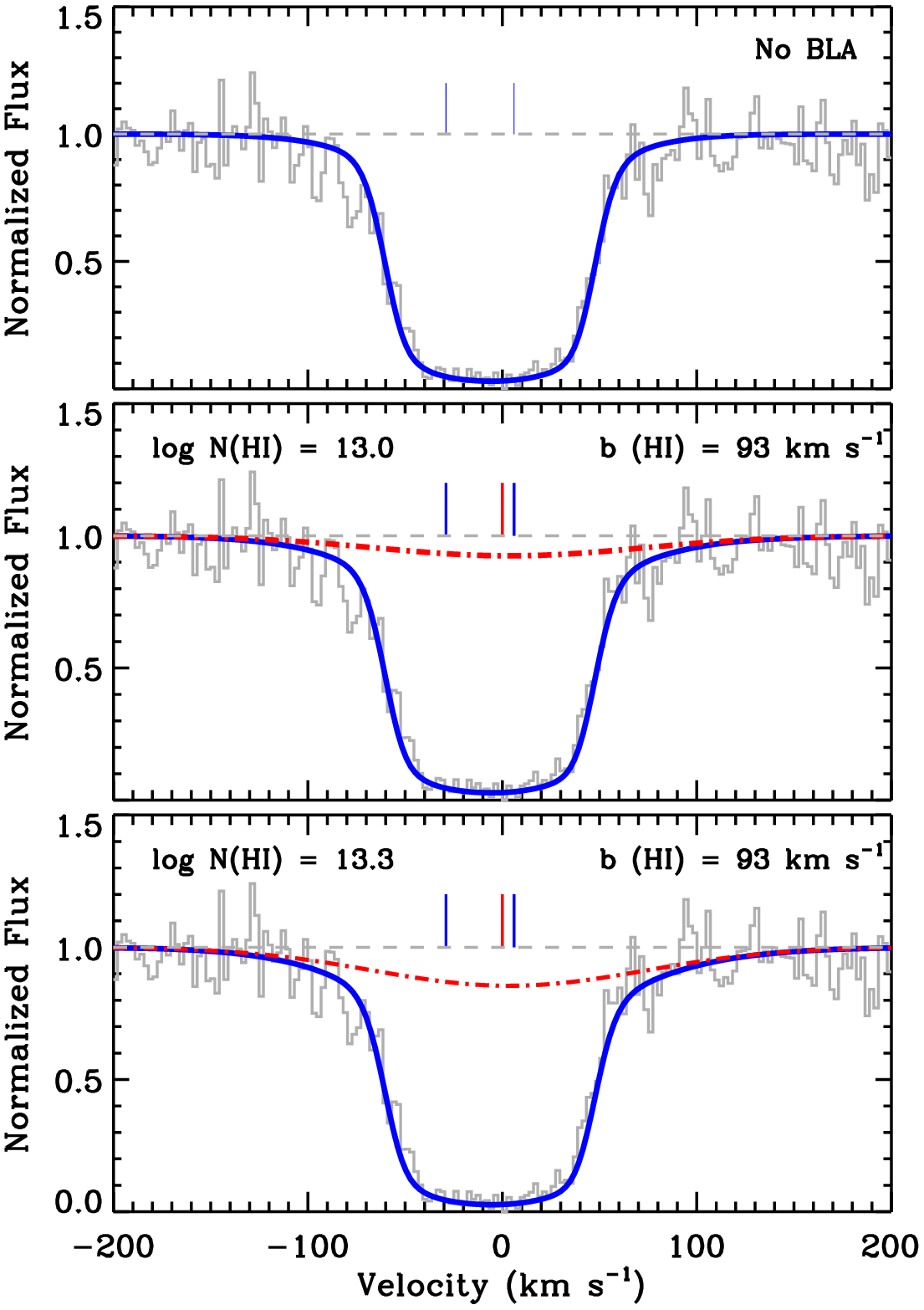}
\end{center}
\protect
\vspace{0.5in}
\caption{\large The {\it top panel} shows the {\Lya} line in the $z = 0.32566$ absorber with the profile model ({\it blue}) superimposed. The model was obtained by simultaneously fitting {\Lya} and six higher order Lyman series lines in the COS spectrum. The best-fit model has two components of approximately equal $N$ and $b$-values at $v = -28~{\pm}~4$~{\kms} and $v = 7~{\pm}~3$~{\kms}. The location of these components are marked by the vertical ticks. The fit parameters are given in Table 3. In the {\it middle} and {\it bottom} panels are shown the cumulative {\Lya} profile obtained when a BLA of $b(\HI) = 93$~{\kms} and $\log [N(\HI)] = 13.0$~dex and $13.3$~dex respectively is added to the absorption at $v = 0$~{\kms}. The BLA profile is shown by the {\it red} curve. The width of the BLA is set by the temperature derived for the measured $N(\NeVIII)/N(\OVI)$ from collisional ionization models. From the cumulative three component model, we can constrain $\log~N(\HI) \sim 13.2$~dex in the BLA.}
\label{fig:6}
\end{figure*}

\end{document}